
\input harvmac

\def\o{\omega}

\Title{USITP-91-18}{ A Novel Chiral Boson}

\centerline{Fiorenzo Bastianelli\footnote{$^\dagger$}{e-mail:
fiorenzo@vana.physto.se}}
\bigskip\centerline
{\it  Institute for Theoretical Physics}
\centerline {\it University of Stockholm}
\centerline {\it Vanadisv\"agen 9 }
\centerline {\it S-113 46 Stockholm, Sweden}
\vskip .6in

\noindent We introduce a new model describing a bosonic system
with chiral properties.
It consists of a free boson  with two peculiar couplings to the
background geometry which
 generalizes the Feigen-Fuchs-Dotsenko-Fateev
construction.
By choosing
 the two background charges of the model, it is possible to achieve
any prefixed value of the left and right central charges and,
in particular, obtain chiral bosonization.
A supersymmetric version of the model is also given. We use
the latter to identify the effective action induced by chiral
superconformal matter.

\Date{11/91}

Bosonization is an interesting phenomenon of two dimensional field theory.
It states the equality of correlation functions of given operators
in the two apparently different theories of a scalar boson
and a Dirac fermion
\ref\Col{S. Coleman, Phys. Rev. {\bf D11} (1975) 2088\semi
S. Mandelstam, Phys. Rev. {\bf D11} (1975) 3026.}.
In the massless case, it has been useful to extend the fermion-boson
correspondence
to a single chiral sector, and
formulas for chiral bosonization on a general Riemann
surface have been written down in
\ref\VV{
E. Verlinde and H. Verlinde, Nucl. Phys. {\bf B288} (1987) 357\semi
T. Eguchi and H. Ooguri, Phys. Lett. {\bf B187} (1987) 127\semi
L. Alvarez-Gaum\'e, J.B. Bost, G. Moore, P. Nelson and C. Vafa,
 Comm. Math. Phys. {\bf 112} (1987) 503.}.
However, the chiral bosonic
piece has always been identified
by
extracting the holomorphic (in the modular parameters)
square root
of an  effective action which
corresponds  to a non-chiral bosonic theory,
and a direct comparison with a truly chiral bosonic theory
 is still absent. The reason is that
it is difficult to give a lagrangian description of a chiral  boson
 suitable for quantization.
The first attempt  to construct such an action was made by Siegel
\ref\War{W. Siegel, Nucl. Phys. {\bf B238} (1984) 307.}, who
introduced a gauge field in order to constrain  the boson to be chiral.
Unfortunately,
the gauge symmetry becomes anomalous upon quantization
and mechanisms to  cancel the anomaly have to  be devised
\ref\Hull{C. Hull, Phys. Lett. {\bf B206} (1988) 235; {\bf B212} (1988) 347. }.
A different formulation,
due to
Floreanini and Jackiw
\ref\FJ{R. Floreanini and R. Jackiw, Phys. Rev. Lett. {\bf 59}
 (1987) 1973.},
 does not require the introduction of
a gauge field, but the  Lorentz invariance of the model is not manifestly
realized.
This is not a problem in principle, in fact one can still
achieve several results like coupling the chiral boson to other systems
\ref\B{F. Bastianelli and P. van Nieuwenhuizen, Phys. Lett.
{\bf B217} (1989).},
computing the gravitational anomalies
\ref\Ba{F. Bastianelli and P. van Nieuwenhuizen, Phys. Rev. Lett.
{\bf 63} (1989)
728.}
and showing chiral bosonization on a curved space with trivial topology
(i.e. the infinite plane)
\ref\Bas{ F. Bastianelli, Phys. Lett. {\bf B254} (1991) 427.}.
However, the non-manifest Lorentz invariance (or coordinate
invariance when  coupled to
gravity) makes it difficult to investigate
the
Floreanini-Jackiw model in
non-trivial topological situations.
In the present letter, we suggest yet a new description of a chiral boson
 which is different in spirit
from the previous ones.
The model we have in mind is that of a free scalar field coupled to
the background geometry
through two background charges.
These couplings induce
 improvement terms in the energy-momentum
 tensor
and modify
 the left and right central charges of the model,
in a way similar to the usual boson with a background charge,
 discussed by Feigin-Fuchs and Dotsenko-Fateev
\ref\ff{B.L. Feigin and D.S. Fuchs, Funct. Anal. Appl. {\bf 16} (1982) 114
\semi Vl.S. Dotsenko and V.A.
Fateev, Nucl.Phys. {\bf B240} [FS12] (1984) 312; {\bf B251} [FS13] (1985)
691.}.
One can choose the value of the two background charges
to create a mismatch between the left and right
central charges, thus obtaining a chiral system.
We present only a local description of the model (more
precisely, we work on the infinite plane) and adopt a
Minkowskian
signature for the two dimensional geometry.
We conclude this letter by presenting
 a supersymmetric version of the chiral boson and its effective
action. The latter is naturally identified with the typical
effective action induced by chiral superconformal matter.

We start with the definition
of the two dimensional geometry and  introduce
 Lorentz covariant derivatives\footnote
{$^\dagger$}{$J$ is the Lorentz generator acting on a vectors $V_\pm$
as follows $[J,V_\pm] = \pm V_\pm$;
the Minkowski metric in light cone coordinates
is  given by
$\eta_{+- }=\eta_{-+}={1\over2}$ and $ \eta_{++} = \eta_{--}=0$.}
\eqn\1{ \nabla_a = E_a{}^\mu \partial_\mu + \o_a J;\ \ \ \ \ \ \ \ \ a=\pm}
constrained by
\eqn\2{ \lbrack \nabla_+, \nabla_- \rbrack = R J}
The constraint \2 can be solved to get
 the usual spin connection $\o_a$ as function of the vielbein field $E_a{}^\mu$
($e \equiv det\  e_\mu{}^a$, $e_\mu{}^a$ is the inverse of $E_a{}^\mu$)
\eqn\3{ \o_\pm = \mp {1\over e} \partial_\mu (e E_\pm^\mu ) }
The scalar curvature $R$ defined in \2 can be expressed as follows
($E_a \equiv E_a{}^\mu \partial_\mu$)
\eqn\R{\eqalign {&R= R_l +R_r\cr
&R_l =\nabla_+ \o_- = (E_+ -\o_+) \o_-\cr
&R_r =-\nabla_- \o_+ = -(E_- +\o_-) \o_+\cr}}
Note that $R_l$ and $R_r$ are not separately local Lorentz invariant.
In fact, under  local Lorentz ($\lambda$) and Weyl ($\sigma$)
transformations defined by
\eqn\lor{ \hat e_\mu{}^\pm = {\rm exp}(\sigma \mp \lambda) e_\mu{}^\pm}
one obtains the following transformed quantities
\eqn\tr{ \eqalign{ &\hat e \hat R_l =
e \biggr ( R_l + \nabla^2 (\sigma -\lambda )\biggl ) \cr
&\hat e \hat R_r = e \biggr
( R_r + \nabla^2 (\sigma +\lambda )\biggl )\cr}}
where $\nabla^2 \equiv \nabla_- \nabla_+ = \nabla_+ \nabla_-$
 acts on scalars. Only the combination
which gives the Ricci scalar $R=R_l +R_r$ is Lorentz invariant.
However, we are looking for a chiral theory where
gravitational anomalies are expected
 \ref\AG{L. Alvarez-Gaum\'e and E. Witten, Nucl. Phys. {\bf B234} (1984)
269.}.
Moreover, it is  known that one
can shift possible  gravitational anomalies into the
Lorentz sector by using local counterterms
\ref\BZ{W. Bardeen and B. Zumino, Nucl. Phys. {\bf B244} (1984) 421.}.
Therefore, it is natural to use independently $R_l$ and $R_r$
in the construction
of a model which is going to be
 reparametrization invariant
but Lorentz (and Weyl) anomalous.

Inspired  by the previous considerations, we write down the following
action for a chiral boson $\phi$
\eqn\act{ S = {1\over  \pi}
\int \ d^2 x e\ \bigl ({1\over2} \nabla_+ \phi \nabla_- \phi +
\beta_r R_r \phi + \beta_l R_l\phi\bigr )}
To justify the claim  that $\phi$ is a chiral boson,
we begin the analysis of the model by looking
at the energy-momentum.
The latter is defined by
$T^\mu{}_a = {2 \pi \over e} {\delta S \over \delta e_\mu{}^a}$
and in general will not be symmetric if the theory is not local
Lorentz invariant
(which is the case of our model).
We obtain
\eqn\str{\eqalign{
T_\mu{}_a &=
e_{\mu a}\nabla_+ \phi \nabla_- \phi
-e_{\mu +}\nabla_a \phi \nabla_- \phi
-e_{\mu -}\nabla_+ \phi \nabla_a \phi
\cr &+2\beta_r \bigl (
e_{\mu a} E_+ \nabla_- \phi
-e_{\mu +} E_a \nabla_- \phi
-e_{\mu -} \o_+ \nabla_a \phi \bigr )
\cr &+2\beta_l \bigl (
e_{\mu a} E_- \nabla_+ \phi
-e_{\mu -} E_a \nabla_+ \phi
+e_{\mu +} \o_- \nabla_a \phi \bigr )
\cr }}
In flat space, where $e_{\mu a} = \eta_{\mu a}$,
we get the following components of the
energy-momentum tensor
\eqn\T{
\eqalign{
& T_{++}= - {1\over 2} \partial_+ \phi \partial_+\phi
- \beta_l \partial_+^2 \phi\cr
& T_{--}= - {1\over 2} \partial_- \phi \partial_-\phi
- \beta_r \partial_-^2 \phi\cr
& T_{+-}= \beta_r \partial_+ \partial_- \phi =0 \cr
& T_{-+} = \beta_l \partial_- \partial_+ \phi =0 \cr}}
where
 we have used  the equation of motion
evaluated in  flat space,
 $\partial_+ \partial_-
\phi =0$, which  also implies the conservation of the energy-momentum
 tensor.
{}From the above expressions we see that the coupling to the
background  geometry has induced
extra terms in the $T_{\pm \pm}$
components.
Their effect is to shift the central charges of the model as follows
\eqn\c{
c_l=1+12\beta^2_l;\ \
c_r=1+12\beta^2_r}
where the central charges $c_l$ and $c_r$ are identified from the coefficients
o
   f the
anomalous terms in the operator product expansion of
the energy-momentum tensor\footnote{$^\dagger$}{
We use the operator product expansion
 $\partial_\pm \phi(x^\pm) \partial_\pm \phi(y^\pm)
= {1\over ( x^\pm -y^\pm)^2} + \cdots $.}
\ref\BPZ{A.A. Belavin, A.M. Polyakov and A.B.
Zamolodchikov,
Nucl. Phys. {\bf B241} (1984) 333.}
\eqn\ope{ \eqalign{
&T_{++} (x^+) T_{++} (y^+) =
 {{c_l/2} \over {(x^+ - y^+)^4}} + {2 T_{++}(y^+)
\over{(x^+ - y^+)^2}} +
{ \partial_+ T_{++}(y^+) \over{(x^+ - y^+)}}
+ \cdots \cr
&T_{--} (x^-) T_{--} (y^-) =
 {{c_r/2} \over {(x^- - y^-)^4}} + {2 T_{--}(y^-)
\over{(x^- - y^-)^2}} +
{ \partial_- T_{--}(y^-) \over{(x^- - y^-)}}
+ \cdots
\cr}}
For $\beta_l^2 \neq \beta_r^2$ we have  a chiral theory and thus a chiral
boson.

An equivalent way of describing the shift
in the central charges is to compute
the effective action in curved space
by integrating out the bosonic field $\phi$ in the
functional integral. This effective action can be interpreted
as the generating functional of correlation functions
of the energy-momentum tensor. We get
\eqn\eff{ \eqalign{
&e^{iW[e_\mu{}^a]} = \int ({\cal D} \phi)
e^{iS}
\cr
&W[e_\mu{}^a] =
{1\over 24 \pi} \int d^2x e \biggl ( R
{1\over \nabla^2}
R +
12 (\beta_l R_l + \beta_r R_r)
{1\over \nabla^2}
(\beta_l R_l + \beta_r R_r) \biggr )
\cr}}
The above result is easily obtained by shifting the field
$\phi \rightarrow \phi + {1\over \nabla^2 }(\beta_l R_l + \beta_r R_r)$
to reduce the computation to that  of a standard  bosonic scalar  without
background charges, which gives the well-known
$R{1\over \nabla^2} R$ effective action
\ref\pol{A.M. Polyakov, Phys. Lett. {\bf B103} (1981) 207.}
with a coefficient of
  $1\over 24 \pi$ in our conventions.
Note that, up to {\it boundary terms}
 (we partially integrate freely
assuming suitable boundary conditions at infinity)
and {\it local terms} (e.g. $ eR_l {1 \over \nabla^2} R_r= e\o_-\o_+  +
$ a total  derivative),
one can rewrite the effective action
as follows
\eqn\effac{
W[e_\mu{}^a] =
{1\over 24 \pi} \int d^2x e
\biggl ( (1 + 12 \beta_l^2 ) R_l {1\over \nabla^2} R_l +
(1 + 12 \beta_r^2 ) R_r {1\over \nabla^2} R_r \biggr )}
which is the effective action of a chiral system, first discussed by
Leutweyler
\ref\Le{H. Leutweyler, Phys. Lett. {\bf B153} (1985) 65.}
for the case of a Weyl fermion
(see also ref.\ref\p{ F. Bastianelli, P. van Nieuwenhuizen and A. Van Proeyen,
Phys. Lett. {\bf B253} (1991) 67.} for useful formulas).
To achieve central charges $c_{r,l} < 1$ one has to take imaginary couplings
$\beta_{r,l}$, as in the non-chiral case \ff. In particular, the choice
$\beta_l =0$ and $\beta_r =\pm {i\over {2 \sqrt 3 }}$ reproduces the
effective action of a Weyl fermion.

It is interesting to observe a relation between the chiral bosonic action
\act \ and the effective action induced by  chiral fields \effac:
the action \act\ is identical to to what may be called the ``Liouville
action'' for the Lorentz mode of the vielbein.
To see this, parametrize the class of Lorentz-equivalent vielbeins
by $e_\mu{}^\pm = {\rm exp}(\mp \lambda) \hat e_\mu{}^\pm$,
where $\hat e_\mu{}^\pm$ is a reference vielbein.
Plugging this parametrization in the action  \effac \  one gets
\eqn\Liou{ W[e_\mu{}^a] = W[\hat e_\mu{}^a] + {1\over 24\pi} \int d^2x \hat e
\biggl \lbrack
(2 + \beta_l^2 +\beta_r^2) \lambda  \hat \nabla^2 \lambda
+ 2 (1 +\beta_r^2) \hat R_r \lambda
- 2 (1 +\beta_l^2) \hat R_l \lambda \biggl \rbrack }
(the inclusion of a local counterterm  of the type
$\Delta S = \int  d^2x e \ \o_- \o_+$
 in the action \effac\
would bring in just some modifications of the various
coefficients).
Thus, we see that  by adjusting the various coupling constants,
the ``Liouville action'' for the Lorentz mode can be used
to describe
a chiral boson.
This is  similar to the  usual Liouville action for the Weyl
mode, which
is used to represent non-chiral matter with arbitrary central charge
(note that  one is typically
interested in $c<1$ and must use imaginary couplings in
the Liouville action).

\def\ppmm{{\pm\pm}}
\def\mmpp{{\mp\mp}}
\def\mm{{--}}
\def\pp{{++}}
A supersymmetric version of the model can  easily be constructed  by
repeating the previous steps in superspace
\ref\H{P.S. Howe,  J. Phys. {\bf A12} (1979)  393.}.
The geometry of the latter is locally
defined by introducing Lorentz covariant
derivatives\footnote{$^\dagger$}{$D_M=\{ D_+, D_-,
\partial_\pp, \partial_\mm \} $
 are the
covariant derivatives of rigid superspace,
where  $ D_\pm = {\partial \over {\partial \theta^\pm}} \pm i \theta^\pm
{\partial \over{\partial x^\ppmm}}$. The Lorentz generator now acts as follows
$[J,V_\pm] = \pm{1\over2} V_\pm$,
$ [J,V_\ppmm] = + V_\ppmm$.}
\eqn\der{ \nabla_A = E_A{}^M D_M + \o_M J; \ \ \ \ \ \ \ \ A=+,-,\pp,\mm }
constrained by
\eqn\con{ \eqalign{
&\{ \nabla_+, \nabla_+ \} = 2i \nabla_{\pp} \cr
&\{ \nabla_-, \nabla_- \} = -2i\nabla_{\mm} \cr
&\{ \nabla_+, \nabla_- \} =  R J \cr}}
The scalar curvature can again be expressed as a sum of chiral pieces
\eqn\sR{ \eqalign{ &R= R_l +R_r\cr
&R_l =\nabla_+ \o_- = \bigl (E_+ -{1\over2}\o_+\bigr ) \o_-\cr
&R_r =\nabla_- \o_+ = \bigl (E_- +{1\over2}\o_-\bigr ) \o_+\cr}}
It is useful to give here the formulas for the fermionic spin connections
\eqn\scon{ \o_\pm = \mp {2\over e} D_M (e E_\pm{}^M)}
from which one can derive the following useful scaling properties
of $R_l$ and $R_r$ under
Weyl ($\Sigma$) and Lorentz ($\Lambda$) transformations
\eqn\sca{\eqalign{&\hat E_\pm{}^M = {\rm exp}(\Sigma \pm \Lambda) E_\pm{}^M \cr
&\hat e \hat R_l = e \biggl ( R_l - 2 \nabla^2 (\Sigma +\Lambda)
\biggr ) \cr
&\hat e \hat R_r = e \biggl ( R_r - 2 \nabla^2 (\Sigma - \Lambda)
\biggr ) \cr}}
where now
$\nabla^2 \equiv \nabla_+ \nabla_- =- \nabla_- \nabla_+ $.
The superspace equivalent of the action \act\ looks similar
\eqn\nact{ S= {1\over 2\pi} \int d^2x d^2\theta e\ ({1\over2}
\nabla_+ \phi \nabla_- \phi + \beta_l R_l \phi + \beta_r R_r \phi)}
and give rise to an affective action identical in form to
\effac \
\eqn\neffac{ W[e_M{}^A] = {1\over 32 \pi} \int d^2x d^2\theta e\
\biggl ( (1 + 8 \beta_l^2 ) R_l {1\over \nabla^2} R_l +
(1 + 8 \beta_r^2 ) R_r {1\over \nabla^2} R_r
+ a \o_+ \o_-
\biggr )}
where the last term represent the ambiguity of adding local
terms consistently with reparametrization invariance (and rigid Lorentz
transformations). This effective action is obtained by
 shifting the field $\phi$ and
by making use of the non-chiral effective action due to a scalar superfield
\ref\p2{A.M. Polyakov, Phys. Lett. {\bf B103} (1981) 211\semi
E. Martinec, Phys. Rev. {\bf D28} (1983) 2604.},
which is still of the type $R {1\over \nabla^2} R$.
It can be seen that
this
effective action
is indeed induced by chiral supermatter.
In fact, one can compute the energy-momentum tensor from the action \nact\
to check that  it corresponds to a typical chiral superconformal system.
The energy-momentum tensor is found by varying the vielbein field
$\delta E_A{}^M \equiv H_A{}^B E_B{}^M$.
One must recall that not all variation $H_A{}^B$ are allowed because
 the constraints \con\ which define the supergeometry must be
preserved.
There are only six independent
variations which can be taken to be $ H_\pm{}^\pm, H_\pm{}^\ppmm,
H_\pm{}^\mmpp $. As for the remaining variations, their
 functional
dependence  is found from varying the
constraints \con.
The energy-momentum tensor
is then identified from
\eqn\ll{\delta S = {1 \over 2\pi}
\int d^2x d^2\theta e\ (-1)^A T_A{}^B H_B{}^A}
where $H_B{}^A$ runs over the above set of independent variations.
In flat space and using the equation of motion, we get the following
non-zero
components of the energy-momentum tensor
\eqn\str{\eqalign{&T_\pp{}^- =
-{1\over 2} D_\pp \phi D_+ \phi - \beta_l D_\pp D_+ \phi \cr
&T_\mm{}^+ =  {1\over 2} D_\mm \phi D_- \phi + \beta_r D_\mm D_- \phi\cr}}
It is straightforward
to check that the above components of the energy-momentum
 tensor generate two copies
of a  superVirasoro
algebra
with central charges $ \hat c_l = 1+8 \beta_l^2 $ and
$ \hat c_r = 1 + 8 \beta_r^2 $
 respectively, where  the central charge $\hat c$ is recognized from
 the operator
product expansion of the spin $\big|{3\over2}\big|$
components of the super energy-momentum
tensor (see \ref\FMS{D. Friedan, E. Martinec and S. Shenker,
Nucl. Phys. {\bf B271} (1986) 93.}
for notations)
\eqn\sst{
T(Z_1) T(Z_2) = {{\hat c /4} \over z_{12}^3}
+{(3/2) \theta_{12} T(Z_2) \over z^2_{12}} +
{(1/2) D_2 T(Z_2) + \theta_{12} \partial_2 T(Z_2) \over z_{12}} +\cdots}
and it is related to the  Virasoro central charge $c$ by $c={3\over2}\hat c$.

To summarize, we have introduced a model of  chiral
bosons which reproduce the correct anomalies of chiral
conformal matter in spaces with trivial topology. We hope that the clear
geometrical formulation will help to discuss the model
also in topologically more interesting situations, like  Riemann surfaces.
The model can easily be supersymmetrized
by working in superspace. The supersymmetric
 version has allowed us to identify
the effective action induced by chiral superconformal matter,
a result which we believe new by itself.

\listrefs
\end